\RequirePackage[2018-12-01]{latexrelease}
\documentclass{rQUF2e}

\usepackage{epstopdf}
\usepackage{subcaption}   
\usepackage{float}
\usepackage{lmodern}
\usepackage{amsfonts}
\usepackage{amssymb}
\usepackage{scalerel}

\usepackage{booktabs}
\usepackage{makecell}
\usepackage{multirow}

\usepackage{enumitem}

\usepackage{xcolor}

\theoremstyle{plain}

\theoremstyle{definition}

\theoremstyle{remark}

\newcommand*\colvec[3][]{
    \begin{pmatrix}\ifx\relax#1\relax\else#1\\\fi#2\\#3\end{pmatrix}
}

\usepackage[hidelinks,
            hypertexnames=false,
            naturalnames=false,
            plainpages=false,
            bookmarksnumbered=true]{hyperref}

\makeatletter
\let\rQUF@orig@newlabel\newlabel
\def\newlabel#1#2{\rQUF@processnewlabel{#1}#2\@nil}
\def\rQUF@processnewlabel#1#2#3#4#5#6\@nil{%
  \rQUF@orig@newlabel{#1}{{#2}{#3}}%
}
\makeatother

\makeatletter
\def\fnum@figure{Figure~\thefigure}
\def\fnum@table{Table~\thetable}
\makeatother
\begin{document}
\title{Multiscaling in the Rough Bergomi Model: A Tale of Tails}
\author{
    G. BRANDI$\dag$
    and T. DI MATTEO$^{\ast}$$\ddag$$\S$$\|$\thanks{$^\ast$Corresponding authors: Email: \href{mailto:tiziana.di_matteo@kcl.ac.uk}{\nolinkurl{tiziana.di_matteo@kcl.ac.uk}}}\\
    \affil{
        $\dag$CoMENS, Northeastern University London, London, UK\\
        $\ddag$Department of Mathematics, King's College London, London, UK\\
        $\S$Complexity Science Hub Vienna, Vienna, Austria\\
        $\|$Centro Ricerche Enrico Fermi, Via Panisperna 89 A, Rome, Italy
    }
    \received{Received XX Month 20XX; in final form XX Month 20XX}
}

\maketitle

\begin{abstract}
The rough Bergomi (rBergomi) model, characterised by its roughness parameter $H$, has been shown to exhibit multiscaling behaviour as $H$ approaches zero. Multiscaling, therefore, has profound implications for financial modelling: it affects extreme risk estimation, influences optimal portfolio allocation across different time horizons, and challenges traditional option pricing approaches that assume uniscaling behaviour. Understanding whether multiscaling arises primarily from the roughness of volatility paths or from the resulting fat-tailed returns has important implications for financial modelling, option pricing, and risk management. This paper investigates the real source of this multiscaling behaviour by introducing a novel two-stage statistical testing procedure. In the first stage, we establish the presence of multiscaling in the rBergomi model against an uniscaling fractional Brownian motion process. We quantify multiscaling by using weighted least squares regression that accounts for heteroscedastic estimation errors across moments. In the second stage, we use shuffled surrogates that preserve return distributions while destroying temporal correlations. This is done by using distance-based permutation tests robust to asymmetric null distributions. To validate our procedure, we assess the robustness of the results using synthetic processes with known multifractal properties, namely the Multifractal Random Walk (MRW) and the Fractional L\'evy Stable Motion (FLSM). We provide compelling evidence that multiscaling in the rBergomi model arises primarily from fat-tailed return distributions rather than memory effects. Our findings suggest that the apparent multiscaling in rough volatility models is largely attributable to distributional properties rather than genuine temporal scaling behaviour.
\end{abstract}


\begin{classcode}
C58; C22; C12; C15;  G12
\end{classcode}

\section{Introduction}
\label{sec:intro}

The advent of rough volatility models has significantly transformed financial modelling in recent years, offering superior fits to both implied volatility surfaces and historical time series data \citep{gatheral2018volatility,bennedsen2017decoupling}. These models, characterised by volatility paths with low regularity, have gained attention for their ability to capture both the stylised facts of asset returns and the empirical features of option markets. The roughness of volatility was initially identified through the analysis of realised volatility time series, with empirical studies consistently finding Hurst parameters well below 0.5 \citep{gatheral2018volatility,fukasawa2017volatility}. A more nuanced picture has emerged from recent empirical work on SPX and VIX option data. \citet{romer2022empirical} finds that, while the rough Bergomi and rough Heston models calibrate to SPX options better than the classical Heston model, they do not generate a sufficiently pronounced term structure of the smile, nor the greater symmetry of short-expiry smiles relative to long-expiry ones; a classical two-factor model driven by Ornstein--Uhlenbeck processes is shown to achieve near-perfect SPX fits, and decent joint SPX--VIX fits, without roughness. \citet{guyon2023ATMskew} document that the term structure of the equity at-the-money skew follows a power law only away from short maturities: the power-law fit deteriorates considerably once short maturities are included, and three-parameter shapes that do not blow up as maturity vanishes fit the data better than the two-parameter power law associated with one-factor rough volatility. \citet{delemotte2023ssrn} reach a compatible conclusion, describing the skew term structure as a crossover between distinct power-law regimes rather than a single power law, which would point towards two rough factors with different Hurst exponents, or a single factor with a maturity-dependent Hurst exponent. Taken together, these studies confirm the empirical relevance of low-regularity volatility dynamics while also delineating the limits of the one-factor rough paradigm, and in particular the adequacy of a single power-law skew governed by one constant $H$. Since the object of the present study is the scaling behaviour of the model's sample paths rather than its option-pricing performance, we retain the one-factor rBergomi specification, which remains the canonical rough volatility model and the one for which the multiscaling question was originally posed \citep{brandi2022multiscaling}. Independent confirmation has also come from machine-learning approaches: \citet{rosenbaum2022universality} show that a model-free LSTM forecaster trained on a pool of liquid stocks recovers the same scaling behaviour as the rough fractional stochastic volatility framework, providing nonparametric evidence for the universality of the rough volatility paradigm. Among these models, the rough Bergomi (rBergomi) model introduced by \citet{bayer2016pricing} has emerged as a prominent framework. It incorporates fractional Brownian motion (fBm) with Hurst parameter $H < 1/2$ to generate rough volatility paths. As the roughness parameter $H$ approaches zero, the volatility process becomes increasingly irregular, better matching the observed characteristics of financial markets. The model's success in capturing market behaviour has been demonstrated through various applications, including option pricing \citep{bayer2018pricing,mccrickerd2017turbo}, VIX derivatives \citep{jacquier2018vix,bonesini2021functional}, and portfolio optimization under rough volatility \citep{abijaber2020markowitz}.

The theoretical foundations of rough volatility have been further developed through connections with market microstructure. \citet{elEuch2018microstructure} and \citet{jusselin2018noarbitrage} demonstrated that rough volatility naturally emerges from Hawkes process models of market activity, providing a microscopic foundation for the observed roughness. More recently, \citet{horvath2024convergence} showed how heavy-tailed Hawkes processes converge to rough volatility models, establishing a direct link between market microstructure and the empirical properties of volatility. The universality of volatility formation has been further explored by \citet{rosenbaum2022universality}, who demonstrated that machine learning approaches independently recover the rough volatility paradigm from market data.

Several studies have explored how rough volatility models affect option pricing, particularly for exotic derivatives. \citet{alos2018smile} investigated the smile properties of volatility derivatives under rough volatility, while \citet{horvath2018volatility} examined the pricing of volatility options in rough volatility models. The impact on implied volatility surfaces has been studied extensively, with \citet{fukasawa2022asymptotically} providing asymptotic expansions and \citet{friz2022diamonds} developing forward variance models that capture the rough volatility dynamics. Recent advances in numerical methods have made the implementation of rough volatility models more practical. \citet{bayer2020weak} developed weak error rates for option pricing under linear rough volatility, while \citet{bourgey2021multilevel} introduced multilevel Monte Carlo methods for the rough Bergomi model. Machine learning approaches have also been applied, with \citet{bayer2018deep} using deep learning for calibration and \citet{jacquier2019random} employing random neural networks for rough volatility modelling.

A particularly intriguing property observed in both empirical studies and theoretical analyses is that the rBergomi model exhibits multiscaling behaviour, especially for small values of $H$ \citep{brandi2022multiscaling,comte2012affine,forde2022riemann}. Multiscaling implies that different moments of price changes scale with different exponents, contradicting the simple scaling found in classical Brownian motion models. This property has been interpreted as evidence of complex dynamics in financial markets that go beyond standard diffusion processes. The presence of multiscaling in financial markets was initially proposed by \citet{mandelbrot1997multifractal} and has since been documented across various asset classes and markets \citep{calvet2002multifractality,di2003scaling,green2014origins,zhou2009components,barunik2012understanding,buonocore2016measuring,kantelhardt2002multifractal,buonocore2020interplay,di2007multi,jiang2019multifractality}.
The results of our investigation have significant implications for financial modelling, option pricing, and risk management. They contribute to the ongoing debate about the nature of financial market complexity and the appropriate mathematical frameworks for capturing this complexity, building on the foundational work of \citet{gatheral2018volatility} and extending it to address the fundamental question of whether rough volatility models capture genuine temporal complexity of asset returns or primarily reflect distributional phenomena.
Indeed, the relationship between rough volatility and multiscaling has important implications for derivative pricing and risk management.

However, despite these theoretical and computational advances, a fundamental question remains open: what is the primary source of the multiscaling behaviour observed in rough volatility models? The literature has identified three potential sources of multiscaling. First, multiscaling can arise solely from fat-tailed (non-Gaussian) distributions, with temporal structure playing no essential role beyond creating these distributional properties \citep{zhou2009components}. Second, multiscaling can emerge purely from complex memory structures and temporal dependencies, even in the absence of fat tails, as demonstrated in long-range correlated Gaussian processes \citep{green2014origins}. Third, multiscaling can result from the combination of both distributional effects and temporal dependencies, where fat tails and memory contribute together to the scaling behaviour \citep{stanley2002physica}. While all three mechanisms are theoretically possible, distinguishing between them requires careful statistical methodology.

The distinction between these hypotheses has profound implications. Recent work by \citet{chong2022statistical} on statistical inference for rough volatility has highlighted the importance of understanding the sources of apparent roughness, while \citet{fukasawa2020volatility} questioned whether volatility truly needs to be rough or whether distributional effects could explain the observed phenomena. The debate extends to the interpretation of market efficiency and the nature of price formation processes, as discussed in \citet{gatheral2020quadratic} in the context of the quadratic rough Heston model.

This paper presents a systematic investigation of this question, using surrogate data methods to isolate the contributions of distributional properties and memory effects to the observed multiscaling in the rBergomi model \citep{zhou2009components}. Through statistical testing, we aim to determine whether the multiscaling observed in rough volatility models is simply an artefact of their distributional characteristics or if it also reflects genuine temporal complexities. Our approach involves two key steps: first, we establish the presence of multiscaling in the rBergomi model using appropriate benchmark processes; second, we apply surrogate data techniques to disentangle distributional and temporal sources of multiscaling.

The remainder of this paper is structured as follows. Section~\ref{sec:rbergomi} introduces the rough Bergomi model and discusses its key properties, with particular emphasis on the role of the roughness parameter $H$. Section~\ref{sec:scaling} presents the generalised Hurst exponent framework for quantifying multiscaling behaviour in financial time series. Section~\ref{sec:source} develops our statistical methodology, including surrogate data generation techniques and hypothesis testing procedures for identifying the source of multiscaling. Section~\ref{sec:results} presents the validation of the methodology on synthetic processes and our main results for the rough Bergomi model across different roughness regimes. Section~\ref{sec:conclusion} concludes.

\section{The Rough Bergomi Model}
\label{sec:rbergomi}

The rough Bergomi (rBergomi) model of \citet{bayer2016pricing} extends the
classical stochastic volatility framework by driving the variance process
with a fractional kernel that produces volatility paths of low regularity.
On a filtered probability space $(\Omega,\mathcal{F},
(\mathcal{F}_t)_{t\in[0,T]},\mathbb{Q})$ the asset price $S(t)$ and the
instantaneous variance $v(t)$ satisfy
\begin{align}
\frac{dS(t)}{S(t)} &= \sqrt{v(t)}\,\Big(\rho\,dW(t)
                       + \sqrt{1-\rho^{2}}\,dW^{\perp}(t)\Big),
\label{eq:rb_S}\\[2pt]
v(t) &= \xi_{0}(t)\,
        \exp\!\left(\eta\,\widehat{W}^{H}(t)
        -\tfrac{1}{2}\,\eta^{2}\,t^{2H}\right),
\qquad t\in[0,T],
\label{eq:rb_v}
\end{align}
where $W$ and $W^{\perp}$ are independent standard Brownian motions, the
parameter $\rho\in[-1,1]$ is the leverage correlation, $\eta>0$ is the
volatility of volatility, and $\xi_{0}(t)=
\mathbb{E}^{\mathbb{Q}}[v(t)\mid\mathcal{F}_{0}]$ is the deterministic
initial forward variance curve. Throughout this paper we adopt a flat
forward variance, $\xi_{0}(t)\equiv\xi_{0}$, in line with the bulk of the
simulation-based rBergomi literature \citep{bayer2016pricing,
mccrickerd2017turbo,bennedsen2017hybrid}.

The fractional driver $\widehat{W}^{H}$ in~\eqref{eq:rb_v} is the
Riemann--Liouville (truncated) fractional process introduced in this
context by \citet{bayer2016pricing}:
\begin{equation}
\widehat{W}^{H}(t) \;=\; \sqrt{2H}\int_{0}^{t}(t-s)^{H-1/2}\,dW(s),
\qquad H\in(0,1/2).
\label{eq:RL_def}
\end{equation}
Note that $\widehat{W}^{H}$ shares its driving Brownian motion $W$ with the
price equation~\eqref{eq:rb_S}: the leverage correlation $\rho$ is
therefore realised through this common $W$, while $W^{\perp}$ supplies the
idiosyncratic component of returns. The process $\widehat{W}^{H}$ is a
Gaussian Volterra process; its sample paths are H\"older continuous of
any order strictly less than $H$, which yields roughness increasing as
$H\to 0$.

It is important to stress that $\widehat{W}^{H}$ is not the standard
two-sided fractional Brownian motion (fBm) with covariance
$\tfrac{1}{2}(t^{2H}+s^{2H}-|t-s|^{2H})$. The Riemann--Liouville and
two-sided fBm processes share the same H\"older regularity and the same
small-time scaling, but their covariance functions differ; in particular,
the covariance of $\widehat{W}^{H}$ involves a hypergeometric correction
and is not stationary in increments (see Section~4 of
\citealp{bayer2016pricing} for an explicit expression). The simulation
methodology described below is consistent with the Riemann--Liouville
representation~\eqref{eq:RL_def}, as is the discussion of theoretical
results in subsequent sections.

\subsection{Simulation methodology}
\label{ssec:simulation}

We simulate paths of \eqref{eq:rb_S}--\eqref{eq:RL_def} using the
\emph{hybrid scheme} of \citet{bennedsen2017hybrid}, which is purpose-built
for Brownian semi-stationary processes of Riemann--Liouville type. The
hybrid scheme combines an exact Wiener-integral treatment of the singular
near-zero region of the kernel $(t-s)^{H-1/2}$ with a discrete Riemann-sum
approximation of the regular far-region, and converges at a strong rate of
$n^{-(H+\kappa)}$ for $\kappa\geq 1$ grid points in the singular window.
We use $\kappa=1$ throughout, which is the standard choice and is well
validated for $H\in(0,1/2)$.

An exact alternative would be to simulate $\widehat{W}^{H}$ via the
Cholesky factorisation of its full $N\times N$ covariance matrix
\citep{Davies1987,bennedsen2017hybrid}. For our resolution and replication
count ($N=10{,}000$ time steps, $R=1000$ Monte Carlo replicates per
parameter value, six values of $H$), exact Cholesky simulation is
computationally prohibitive: the cost is $O(N^{2})$ in memory and
$O(N^{3})$ per realisation. The hybrid scheme, by contrast, runs in
$O(N\log N)$ time per realisation via Fast Fourier Transform and has been
extensively validated against Cholesky and other exact methods, including
in the limit $H\to 0$ where bias concerns are sharpest
\citep{bennedsen2017hybrid,mccrickerd2017turbo}. Weak error rates for
rBergomi option pricing under the hybrid scheme have been established by
\citet{bayer2020weak} and refined by Gassiat (2023), confirming that the
scheme is accurate at the resolutions we employ. Independently, our
results in the rough regime $H\to 0$ are consistent with the theoretical
small-$H$ limits derived by \citet{forde2022riemann}, providing a further
sanity check on the simulation.

\subsection{Choice of parameters}
\label{ssec:params}

For our analysis we generate sample paths with different values of $H$ in
the range $[0.001,0.30]$ while fixing the remaining parameters at the
empirically relevant levels $\rho=-0.9$ and $\eta=1.9$, in line with
calibrations to SPX option data reported by
\citet{bayer2016pricing,gatheral2018volatility} and subsequent works.
We use an initial forward variance of $\xi_{0}=0.1$.\footnote{Practitioner
calibrations to SPX data typically report somewhat lower values around
$\xi_{0}\approx 0.04$ \citep{bayer2016pricing}. The choice of $\xi_{0}$
is, however, immaterial for the scaling analysis that follows: $\xi_{0}$
enters the variance process~\eqref{eq:rb_v} multiplicatively, so under
$\xi_{0}\to c\,\xi_{0}$ the structure functions transform as
$\Xi(\tau,q)\to c^{q/2}\,\Xi(\tau,q)$, i.e.~a constant additive offset in
$\log\Xi(\tau,q)$ that affects only the intercept of the log--log
regression. The slope $H(q)$, the multiscaling proxy $B$, the kurtosis
of returns, and all autocorrelation-based statistics are therefore
invariant under $\xi_{0}$-rescaling.}

\paragraph{Empirical estimates of $H$}
Empirical estimates of the roughness parameter $H$ in financial markets
typically fall in the range $0.05$--$0.15$, well below the Brownian
value $0.5$ \citep{gatheral2018volatility,bennedsen2017decoupling}.
\citet{rosenbaum2022universality} provide independent evidence in favour
of this range via a model-free machine-learning approach. Estimates of this order are also
obtained in recent option-based calibration studies, although these
studies emphasise that $H$ is not sharply identified:
\citet{guyon2023ATMskew} and \citet{delemotte2023ssrn} show that the
implied-skew term structure is not well described by a single power law
across maturities, so that the implied $H$ depends on the maturity range
used, while \citet{romer2022empirical} obtains comparable option fits
with non-rough two-factor dynamics. We treat $H$ accordingly as a
control parameter to be varied over a grid, rather than as a quantity to
be calibrated to a single value. We therefore focus
our simulation grid on the range $[0.001,0.30]$, with particular attention
to the empirically relevant subrange $[0.05,0.20]$.

\subsection{Effect of the roughness parameter}
\label{ssec:effects}

\begin{figure}[H]
    \centering
\includegraphics[width=0.9\textwidth]{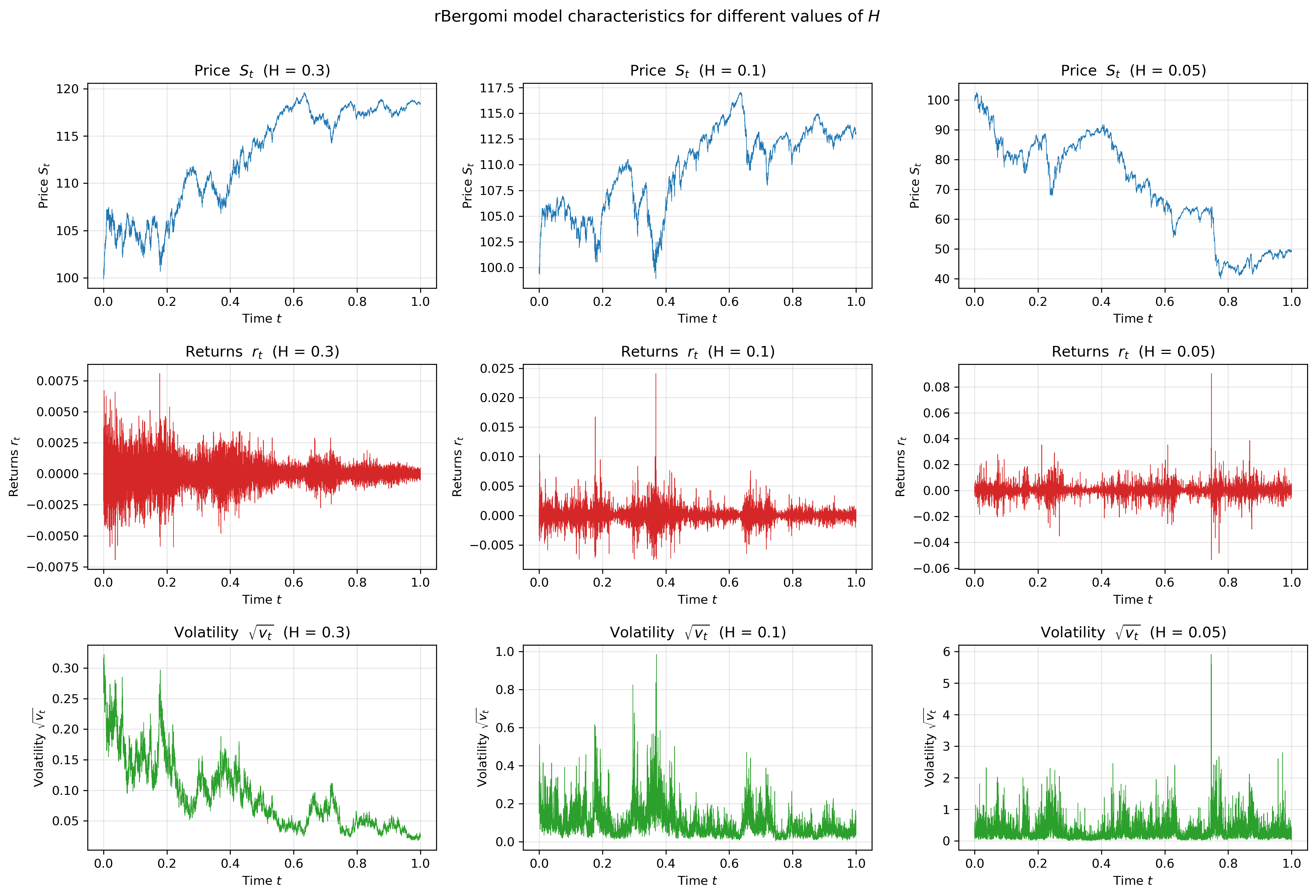}
    \caption{Rough Bergomi model characteristics for three values of $H$.
    Top row: simulated price paths $S(t)$ over $N=10{,}000$ steps.
    Middle row: log-returns $r_{t}=\log\big(S(t)/S(t-1)\big)$ from the
    same paths. Bottom row: spot volatility $\sqrt{v(t)}$ from the
    variance process. Parameters: $\xi_{0}=0.1$, $\rho=-0.9$,
    $\eta=1.9$, generated via the hybrid scheme of
    \citet{bennedsen2017hybrid} with $\kappa=1$. The same random seed is
    used across columns. As $H$ decreases, paths exhibit larger
    and more clustered excursions, and volatility exhibits more
    pronounced clustering and roughness.}
    \label{fig:rbm_characteristics}
\end{figure}

Figure~\ref{fig:rbm_characteristics} illustrates the effect of $H$ on
simulated paths from \eqref{eq:rb_S}--\eqref{eq:RL_def}. Three values of
$H$ are shown ($H=0.3,\,0.1,\,0.05$), generated with a common random seed
and presented with fixed y-axes within each row to highlight the
qualitative changes induced by $H$ rather than rescaling effects.

Three effects are visible. First, price paths become increasingly
irregular and exhibit larger, more clustered excursions; we emphasise
that the model is path-continuous and these are not jumps in the
probabilistic sense, but rather rapid continuous movements driven by the
roughness of the variance. Second, the log-returns develop substantially
heavier tails as $H$ decreases. Third, the spot volatility process
exhibits stronger clustering and visibly rougher sample paths as $H$
decreases.

\begin{figure}[H]
    \centering
\includegraphics[width=0.95\textwidth,height=0.5\textheight]{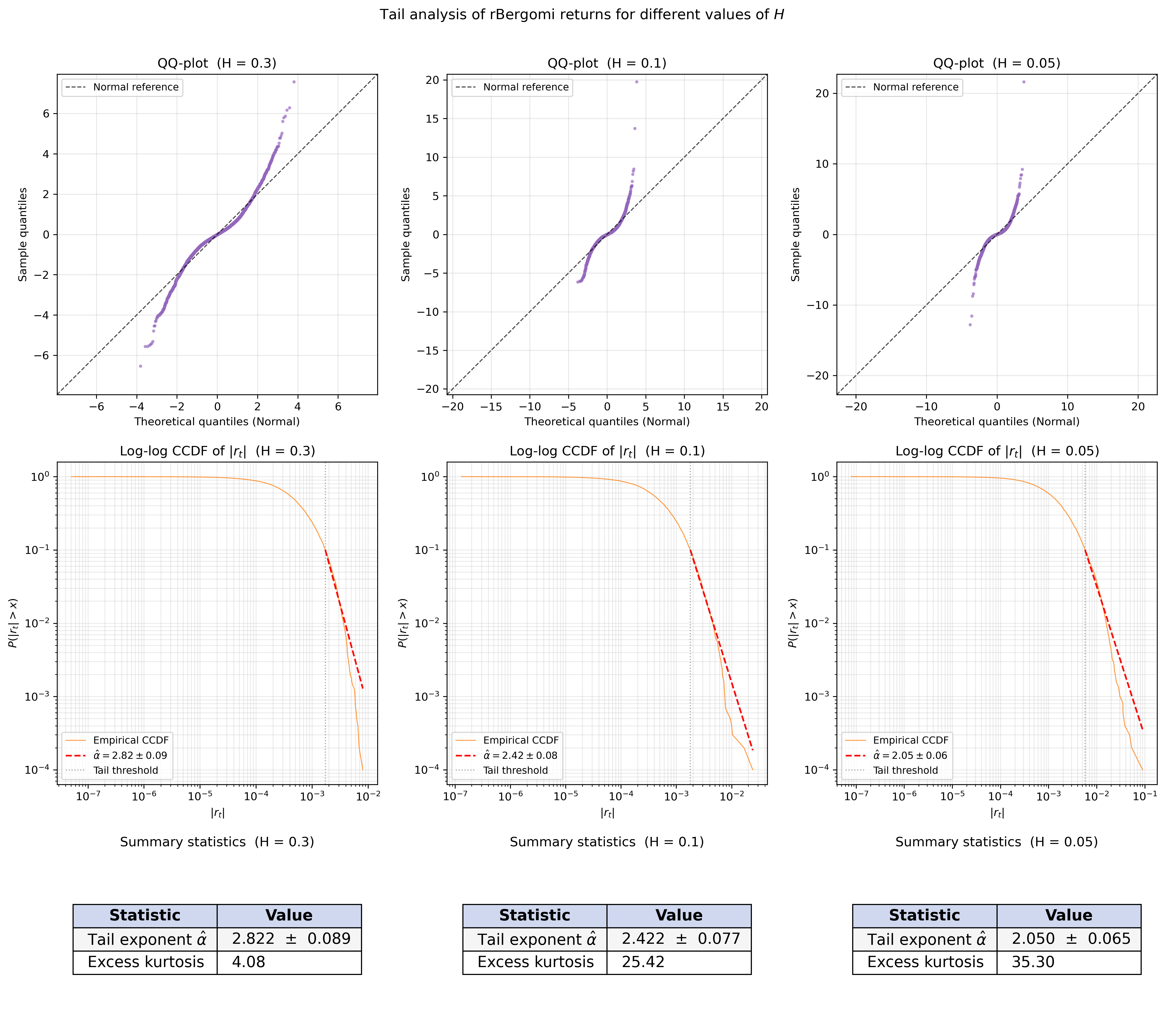}
    \caption{Tail analysis of rBergomi returns for three values of the
roughness parameter $H$. Top row: QQ-plots of the log-returns
$r_{t}$ against the standard normal distribution; the dashed line
is the diagonal reference. Middle row: log--log plot of the
empirical CCDF $P(|r_{t}|>x)$ of absolute returns, with a power-law
fit (dashed red line) above the tail threshold (dotted vertical
line); the legend reports the estimated tail exponent
$\widehat{\alpha}$ with its standard error. Bottom row: tabulated
summary statistics for each panel, including $\widehat{\alpha}$
with its standard error and the excess kurtosis. As $H$ decreases
the return distribution becomes increasingly heavy-tailed: the
estimated tail exponent declines from $2.82$ at $H=0.3$ to $2.05$
at $H=0.05$, and the excess kurtosis grows from $4.08$ to $35.30$.
Simulation parameters are as in Figure~\protect\ref{fig:rbm_characteristics}.}
    \label{fig:rbm_tails}
\end{figure}

To characterise the heavy-tailed behaviour of the returns more
precisely, Figure~\ref{fig:rbm_tails} reports a dedicated tail analysis
at the same three values of $H$. The top row shows the QQ-plots of the
log-returns against the standard normal distribution: at $H=0.3$ the
sample quantiles align with the reference diagonal up to about $\pm 3$
standard deviations, while at $H=0.1$ they already reach $\pm 20$ and at
$H=0.05$ the upper tail extends beyond this range, indicating extreme
tail behaviour. The middle row shows the empirical complementary
cumulative distribution function (CCDF) of $|r_{t}|$ on a log--log
scale, with a power-law fit to the tail. We estimate the tail exponent
$\widehat{\alpha}$ via the \citet{hill1975} maximum-likelihood estimator
applied to the top $10\%$ of $|r_{t}|$, and report its asymptotic
standard error $\widehat{\text{SE}}(\widehat{\alpha})=
\widehat{\alpha}/\sqrt{k}$, where $k$ is the number of tail
observations \citep{embrechts1997}. This sharp transition in tail
heaviness is the primary distributional feature that drives the
multiscaling analysis in the remainder of the paper.

The remainder of the paper investigates the relationship between this
heavy-tailed return distribution and the multiscaling behaviour of the
price process, distinguishing the role played by distributional
properties from that played by temporal dependencies.

\section{Scaling and Multiscaling Analysis in Finance}
\label{sec:scaling}

Multiscaling in financial time series has several important implications
for practical financial applications. The nonlinear scaling of moments
affects the estimation of extreme risks, particularly at longer time
horizons \citep{brandi2022statistics}; traditional risk measures that
assume simple scaling can underestimate tail risks. In portfolio
management, multiscaling affects the relationship between short-term and
long-term investment strategies and influences optimal allocation across
horizons \citep{calvet2002multifractality}. For option pricing, models
that incorporate multiscaling can better capture the term structure of
implied volatility, especially for long-dated options
\citep{bacry2001multifractal}. The presence of multiscaling has also been
interpreted as evidence of market structures that cannot be captured by
standard diffusion models \citep{lux2008markov}. Multiscaling behaviour
has been documented in empirical financial time series across markets and
instruments \citep{di2007multi,jiang2019multifractality}; whether this
multiscaling arises from temporal dependencies or from the heavy-tailed
nature of return distributions remains debated. Throughout this paper we use the standard convention $X(t)$ for a
continuous-time stochastic process and $X_{t}$ for its discrete
observations. When analysing general time series we denote the process by
$X$; when referring specifically to the asset price in the rough Bergomi
model we use $S$. In our numerical analysis the methods developed in the
general setting are applied to the rough Bergomi log-price process, i.e.
$X(t)=\log S(t)$.

\subsection{Generalised Hurst Exponent (GHE) Framework}
\label{ssec:GHE}

For a process $X(t)$ with stationary increments $r_{\tau}(t)=
X(t+\tau)-X(t)$, the GHE methodology characterises the scaling of the
$q$-th absolute moment of the increments with the time-aggregation $\tau$
\citep{di2007multi}:
\begin{equation}
\Xi(\tau,q)\;\equiv\;\mathbb{E}\!\left[|r_{\tau}(t)|^{q}\right]
\;\sim\;K(q)\,\tau^{\,\zeta(q)},\qquad \zeta(q)=q\,H(q),
\label{eq:scaling}
\end{equation}
for $\tau\in[\tau_{\min},\tau_{\max}]$, where $H(q)$ is the generalised
Hurst exponent and $K(q)$ a process-dependent constant. Two broad
regimes can be distinguished. For \emph{uniscaling} processes (e.g.~
Brownian motion, fractional Brownian motion), $H(q)$ is constant in $q$;
for \emph{multiscaling} processes, $H(q)$ varies with $q$, reflecting
intermittent and heterogeneous behaviour. The latter scenario was
proposed for financial markets by \citet{mandelbrot1997multifractal} and
has since been documented across a wide variety of assets
\citep{calvet2002multifractality,di2007multi,gatheral2018volatility}; in
particular, \citet{gatheral2018volatility} report empirical estimates of
$H$ that are remarkably consistent across asset classes.

Taking logarithms of \eqref{eq:scaling},
\begin{equation}
\log\Xi(\tau,q)\;\sim\;qH(q)\log\tau+\log K(q),
\label{eq:loglog}
\end{equation}
which suggests estimating $H(q)$ by linear regression in log--log space. For an observed sample $\{X_{t}\}_{t=0}^{N}$, we estimate the moments by
\begin{equation}
\widehat{\Xi}(\tau,q)\;=\;\frac{1}{N_{\tau}}
\sum_{i=1}^{N_{\tau}}\big|X_{(i+1)\tau}-X_{i\tau}\big|^{q},
\label{eq:Xihat}
\end{equation}
where $N_{\tau}=\lfloor N/\tau\rfloor$ counts non-overlapping increments
of length $\tau$. We then apply two transformations to improve the
numerical conditioning of the subsequent regression. First, we
\emph{normalise} the moments by their value at $\tau=1$,
\begin{equation}
\widetilde{\Xi}(\tau,q)\;=\;\widehat{\Xi}(\tau,q)\big/\widehat{\Xi}(1,q),
\label{eq:Xinorm}
\end{equation}
which eliminates the constant $K(q)$ in~\eqref{eq:scaling}. Second, we
\emph{standardise} by taking the $q$-th root,
\begin{equation}
\dddot{\Xi}(\tau,q)\;=\;\widetilde{\Xi}(\tau,q)^{1/q}\;\sim\;\tau^{H(q)},
\label{eq:Xistand}
\end{equation}
so that all moment orders are placed on the same scale.

Under the transformations \eqref{eq:Xinorm}--\eqref{eq:Xistand}, the
theoretical intercept of the log--log regression is approximately zero,
and we therefore use a no-intercept linear model,
\begin{equation}
\log\dddot{\Xi}(\tau,q)\;=\;H(q)\log\tau+\varepsilon_{\tau,q},
\label{eq:Hreg}
\end{equation}
with $\varepsilon_{\tau,q}$ a mean-zero regression residual. The
estimator of $H(q)$ is the corresponding ordinary least-squares slope,
which we denote $\widehat{H}_{n}(q)$, with standard error
\begin{equation}
\widehat{\sigma}_{H(q)}\;=\;\sqrt{\frac{\sum_{i}
(\,y_{i}-\widehat{y}_{i}\,)^{2}/(n-2)}{q^{2}\sum_{i}(\,x_{i}-\bar{x}\,)^{2}}},
\label{eq:sigmaH}
\end{equation}
where $x_{i}=\log\tau_{i}$, $y_{i}=\log\dddot{\Xi}(\tau_{i},q)$,
$\widehat{y}_{i}$ are the fitted values, and $n=|\mathcal{T}|$ is the
cardinality of the set of scales $\mathcal{T}=\{\tau_{1},\dots,\tau_{n}\}$
used in the regression.

\subsubsection*{Multiscaling proxy}

To quantify the degree of multiscaling we model $\widehat{H}_{n}(q)$ as
an affine function of $q$, following
\citet{brandi2022statistics,brandi2022multiscaling}:
\begin{equation}
H(q)\;=\;A+Bq.
\label{eq:linear_Hq}
\end{equation}
The coefficient $B$ is our \emph{multiscaling proxy}: $B=0$ for
uniscaling processes and $B<0$ for multiscaling processes, with more
negative values indicating stronger multiscaling. Because individual
estimates $\widehat{H}_{n}(q)$ have variances
$\widehat{\sigma}^{2}_{H(q)}$ that grow with $q$ in heavy-tailed
processes, ordinary least squares assigns disproportionate weight to the
noisiest moment estimates. We therefore estimate $(A,B)$ by weighted
least squares,
\begin{equation}
\big(\widehat{A},\widehat{B}\big)\;=\;
\arg\min_{(A,B)}\sum_{k}w_{k}\,
\big[\widehat{H}_{n}(q_{k})-(A+Bq_{k})\big]^{2},
\qquad w_{k}=1/\widehat{\sigma}^{2}_{H(q_{k})},
\label{eq:WLS}
\end{equation}
with standard error of $\widehat{B}$ given by
\begin{equation}
\widehat{\sigma}_{B}\;=\;
\sqrt{\frac{\sum_{k}w_{k}}
{\sum_{k}w_{k}\sum_{k}w_{k}q_{k}^{2}-\big(\sum_{k}w_{k}q_{k}\big)^{2}}}.
\label{eq:sigmaB}
\end{equation}  
We adopt WLS here on principled statistical grounds: $\widehat{\sigma}^{2}_{H(q)}$ grows
steeply with $q$ in heavy-tailed processes, so OLS would be dominated by
the noisiest high-$q$ estimates and would weight all moments equally
regardless of their precision. WLS instead reweights each moment by its
estimated inverse variance, which is the textbook efficient estimator
under heteroscedasticity.\footnote{For processes with mild heteroscedasticity, OLS achieves comparable precision with simpler inference.} The benefit of this choice will be larger on
real financial data, where the heteroscedastic structure across moments
is more pronounced than in the controlled simulations of this paper.

\subsection{Hyperparameter tuning}
\label{ssec:hyper}

Two hyperparameters of the estimation pipeline require careful choice:
the range of moment orders $q$ used in \eqref{eq:linear_Hq}, and the
range of time scales $\tau$ used in the log--log regression
\eqref{eq:Hreg}.

\subsubsection*{Range of moment orders}

For a process with power-law tails of index $\alpha$, the absolute
moments $\mathbb{E}[|X|^{q}]$ exist only for $q<\alpha$
\citep{jiang2019multifractality}. Using $q\geq\alpha$ leads to divergent
moments that, in finite samples, produce unreliable estimates dominated
by rare extreme events; this typically generates a downward bias in
$\widehat{H}(q)$ that mimics multiscaling regardless of the underlying
temporal structure \citep{Barunik2010,heyde2008}. The consequence is that
strong curvature of $\widehat{H}(q)$ observed only at high $q$ is
typically a finite-sample artefact rather than evidence of genuine
multiscaling. Conversely, for a process with genuine multiscaling, the
curvature of $H(q)$ should already be visible at moment orders for which
the moments are finite, including small $q$.

\citet{brandi2022statistics} addressed this issue by restricting the
analysis to $q\leq 1$, motivated by empirical tail-exponent estimates for
financial returns in the range $1.5$--$3$ \citep{Weron2001,Scalas2006,
Eom2019,jiang2019multifractality}. While conservative, this fixed cutoff
is overly restrictive for series with lighter tails and overly liberal
for very heavy-tailed series. In this work we adopt a data-driven
approach that estimates the tail behaviour of each series directly. We
use maximum-likelihood estimation of an $\alpha$-stable distribution
\citep{Nolan2001} fitted to the returns via the fast L\'evy
(quantile-based) estimator; the $\alpha$-stable family is used here not
as a distributional assumption on the returns but as a parsimonious tool
for estimating their tail behaviour. To guard against residual
misspecification we apply a multiplicative safety factor to the estimated
stability parameter,
\begin{equation}
\alpha_{\text{safe}}\;=\;s\cdot\widehat{\alpha}_{\text{stable}},
\qquad s=0.8,
\label{eq:alpha_safe}
\end{equation}
and restrict the analysis to $q_{\max}=\alpha_{\text{safe}}$. Three
considerations motivate this choice. Theoretically, the
maximum-likelihood estimator $\widehat{\alpha}_{\text{stable}}$ has
finite-sample standard error of order $1/\sqrt{N}$
\citep{Nolan2001}, which for $N=10{,}000$ corresponds to a few percent of
the true value; a multiplicative shrinkage of $20\%$ therefore places
$q_{\max}$ several standard errors below the point estimate and ensures
that finite-sample noise in $\widehat{\alpha}_{\text{stable}}$ does not
push $q_{\max}$ above the true tail index. Methodologically, the
choice $s=0.8$ is the conservative value introduced in our earlier
methodological work \citep{brandi2022multiscaling}; using it here
provides consistency with our previous studies and makes results directly
comparable across the two papers. Statistically, the safety
factor is conservative with respect to detecting multiscaling: a
smaller $s$ shrinks the regression range in $q$ and reduces the curvature
of $\widehat{H}_{n}(q)$ that the regression can pick up, making
$\widehat{B}$ closer to zero. The choice $s=0.8$ thus tilts the
methodology against detecting multiscaling, not towards it; the
strong multiscaling signal we report in the very-rough regime ($H\leq
0.01$, multiscaling detected in $100\%$ of simulations) is therefore
robust to plausible variation of $s$ in either direction. We return to
this point in the discussion of the rBergomi results in
Section~\ref{sssec:rb_presence}.

\subsubsection*{Range of time scales}

Several approaches to the choice of $\tau_{\max}$ have been proposed.
\citet{Yue2017} use segmented regression on the structure function
itself; \citet{Buonocore2017} use autocorrelation significance tests;
\citet{brandi2022statistics} introduced the Autocorrelation Segmented
Regression (ACSR) method, which performs segmented regression on the
autocorrelation function of absolute returns. ACSR identifies the scale
at which temporal correlations decay but does not directly ensure that
the scaling relation~\eqref{eq:scaling} holds with good linear fit
quality across all moments $q$.

We adopt a complementary, fit-quality-driven approach. For each candidate
$\tau_{\max}$, we compute the goodness-of-fit $R^{2}(q,\tau_{\max})$ of
the regression~\eqref{eq:Hreg} for each $q\in[q_{\min},q_{\max}]$, and
let
\begin{equation}
R^{2}_{\min}(\tau_{\max})\;=\;\min_{q\in[q_{\min},q_{\max}]}R^{2}(q,\tau_{\max}).
\end{equation}
We select the largest $\tau_{\max}$ for which
$R^{2}_{\min}(\tau_{\max})$ exceeds a stringent threshold,
\begin{equation}
\tau_{\max}^{*}\;=\;\max\bigl\{\tau_{\max}:\;R^{2}_{\min}(\tau_{\max})
\geq 0.95\bigr\}.
\label{eq:taustar}
\end{equation}
This choice ensures that all moments exhibit good power-law scaling over
a common range and that comparisons of $H(q)$ across $q$ values are
based on equally reliable estimates. The threshold $0.95$ is the one
used throughout this paper; it is sharp enough to enforce clean scaling
but not so sharp as to truncate the regression range excessively. Our
method differs from ACSR in that it directly targets the quality of the
scaling relationship rather than inferring it from autocorrelation
structure, which is important when different moments have different
sensitivities to finite-size effects \citep{kantelhardt2002multifractal}.

\section{Source of Multiscaling}
\label{sec:source}

To investigate the source of multiscaling in the rough Bergomi model we
develop a two-stage testing methodology. The first stage tests whether
multiscaling is significantly present at all, against the null of a
matched uniscaling fractional Brownian motion. If multiscaling is
detected, the second stage tests whether it can be reproduced by
preserving the marginal distribution of returns while destroying
temporal dependence, a regime we describe as distributionally
attributable, or whether the observed multiscaling proxy is
significantly different from what shuffled surrogates produce, indicating
that temporal dependence makes a detectable contribution. Our shuffling
procedure preserves the exact return distribution while destroying
temporal correlations, allowing us to isolate the distributional
component of the scaling behaviour.

The approach combines established techniques from multiscaling analysis
with specialised surrogate-data methods designed to isolate the two
candidate sources of scaling, formulated within a hypothesis-testing
framework. Throughout this section we use the symbol
$\mathcal{H}^{(n)}_{\phi}$ to denote a statistical hypothesis, with
$n\in\{1,2\}$ indicating the stage and $\phi\in\{0,A\}$ the null or
alternative; this notation avoids collision with the Hurst exponent $H$.

\subsection{Surrogate data generation}
\label{ssec:surr}

We employ two surrogate-data methods, each isolating a different aspect
of the original series.

\subsubsection{Matched fractional Brownian motion}
\label{sssec:fbm}

For the Stage 1 test we generate fractional Brownian motion (fBm) with
Hurst exponent matched to the first-moment exponent
$\widehat{H}_{n}(1)$ of the original series, using the exact
Davies--Harte method \citep{Davies1987,beran2013}. The Davies--Harte
algorithm simulates fractional Gaussian noise (fGn) via the circulant
embedding of its theoretical autocovariance
$\gamma(k)=\tfrac{\nu^{2}}{2}(|k+1|^{2H}-2|k|^{2H}+|k-1|^{2H})$ and the
diagonalisation of the circulant matrix via the Fast Fourier Transform;
the cumulative sum of the resulting fGn yields a path of fBm. The
resulting process is exactly uniscaling, with population
$H(q)\equiv H(1)$ for all $q$, and reproduces the first-moment scaling
of the data by construction. This is the appropriate null process for
the Stage 1 test, as discussed below.

\subsubsection{Shuffled surrogates}
\label{sssec:shuf}

For the Stage 2 test we generate \emph{shuffled surrogates} by randomly
permuting the returns of the original series
\citep{Theiler1992,Schreiber2000}. For a price series $\{X_{t}\}_{t=0}^{N}$
we form returns $r_{t}=X_{t}-X_{t-1}$, draw a uniformly random
permutation $\pi$ via the Fisher--Yates algorithm
\citep{fisher1938statistical,knuth1998art}, and reconstruct
$\widetilde{X}_{t}=X_{0}+\sum_{s=1}^{t}r_{\pi(s)}$. By construction, this
procedure preserves the exact marginal distribution of returns,
including all moments and tail behaviour, while destroying temporal
dependence at every lag (autocorrelations, volatility clustering,
higher-order dependence).

\subsection{Statistical hypothesis testing framework}
\label{ssec:testing}

Previous approaches to testing for multiscaling have typically examined
whether the estimated slope $\widehat{B}$ in
\eqref{eq:linear_Hq}--\eqref{eq:WLS} differs significantly from zero, via
a standard $t$-test. This approach is unsatisfactory in finite samples:
genuinely uniscaling processes such as fBm produce non-zero values of
$\widehat{B}$ with non-trivial variance due to estimation noise alone.
Testing against the point null $B=0$ ignores this finite-sample
distribution and is liable to over-rejection. We propose a two-stage surrogate-data framework. The first stage tests
for the presence of multiscaling by comparing the observed
$\widehat{B}$ against the empirical distribution of $\widehat{B}$ values
obtained from matched-fBm surrogates, capturing the actual range of
values expected under uniscaling. The second stage, conditional on
detection, uses shuffled surrogates and a distance-based statistic that
remains valid under possibly asymmetric null distributions to test
whether temporal dependence contributes to the observed multiscaling
beyond the marginal distribution of returns.

\subsubsection{Stage 1: presence of multiscaling}
\label{sssec:stage1}

We test the null hypothesis that the data are generated by uniscaling
fBm with the same first-moment scaling as the observed series. Formally,
\begin{equation}
\mathcal{H}^{(1)}_{0}:\quad
\text{the process is uniscaling fBm with } H(q)\equiv H(1).
\end{equation}
The alternative is
\begin{equation}
\mathcal{H}^{(1)}_{A}:\quad
\text{the process exhibits scaling behaviour inconsistent with
uniscaling fBm.}
\end{equation}

The choice of fBm as the benchmark deserves comment. fBm is the
canonical \emph{uniscaling} Gaussian process with a free Hurst exponent
that can be matched to $\widehat{H}_{n}(1)$ of the data. Importantly,
this matching makes the benchmark \emph{less} restrictive rather than
more: any fixed-Hurst null (white noise with $H\approx 0$, standard
Brownian motion with $H=1/2$, or fGn at any other specific value)
would impose its own first-moment scaling on the surrogate
distribution, which need not coincide with the data's. A mismatch in
$H(1)$ between data and benchmark would shift the surrogate
distribution of $\widehat{B}$ for reasons unrelated to multiscaling
and would conflate first-moment scaling with higher-order scaling in
the test statistic. Matching the surrogate Hurst exponent to
$\widehat{H}_{n}(1)$ removes this confound: by construction the
benchmark agrees with the data at the first moment, so any deviation
of $\widehat{B}_{\text{rBergomi}}$ from the surrogate distribution is
attributable to higher-order scaling behaviour. The matched-fBm null
is therefore the most general single-parameter uniscaling benchmark
available, in the sense that it adapts to whatever first-moment
scaling the data exhibit.

Rejecting $\mathcal{H}^{(1)}_{0}$ then means that no uniscaling
fBm with the same first-moment scaling can reproduce the observed
multiscaling proxy. The alternative as stated is therefore
non-parametric: the data are not uniscaling fBm at $\widehat{H}_{n}(1)$,
which we read as evidence of higher-order scaling behaviour. In the
context of financial time series the operational reading of this
alternative is that $B<0$ (high-order moments scale more slowly than
first-order moments), and we therefore implement the test in a
one-sided fashion. We do not claim that rejection of
$\mathcal{H}^{(1)}_{0}$ proves the presence of ``multifractal'' scaling
in the strong cascade-based sense; we claim only that the data are
inconsistent with the matched-fBm uniscaling benchmark.

We generate $M=1000$ independent fBm surrogates with
$H=\widehat{H}_{n}(1)_{\text{rBergomi}}$. For each surrogate $m$ we
compute the multiscaling proxy $\widehat{B}_{\text{fBm},m}$ using
exactly the same estimation pipeline as for the original series, the
same adaptive choice of $q_{\max}$ and $\tau_{\max}$, the same WLS
regression, so that any finite-sample bias and noise are inherited by
the surrogate distribution. The one-sided $p$-value is the empirical
left tail probability,
\begin{equation}
p_{\text{presence}}\;=\;\frac{1}{M}\sum_{m=1}^{M}
\mathbb{I}\bigl(\widehat{B}_{\text{fBm},m}\leq
\widehat{B}_{\text{rBergomi}}\bigr).
\end{equation}
We reject $\mathcal{H}^{(1)}_{0}$ at level $\alpha=0.05$ if
$p_{\text{presence}}<\alpha$.\footnote{An equivalent presentation
standardises the test statistic by the surrogate spread:
$T_{\text{rBergomi}}=(\widehat{B}_{\text{rBergomi}}-\overline{B}_{\text{fBm}})/
\widehat{\mathrm{SD}}(\widehat{B}_{\text{fBm}})$ and similarly for each
surrogate. Since $T$ is a strictly monotone transformation of
$\widehat{B}_{\text{rBergomi}}$ on the support of the surrogate
distribution, the empirical percentile rank, and therefore the
$p$-value, is identical under either statistic. The standardisation
expresses the test statistic in units of standard deviations and aids
interpretation but does not affect inference.}

A natural concern is that the matched-fBm null uses the plug-in
$\widehat{H}_{n}(1)$ in place of the true $H(1)$. Because every
surrogate goes through the same estimation pipeline as the data,
inheriting the same finite-sample bias and noise in $\widehat{H}_{n}(1)$,
the resulting null distribution of $\widehat{B}$ already absorbs
plug-in noise. Moreover, $\widehat{B}$ is by construction the
slope of $\widehat{H}_{n}(q)$ against $q$ rather than its
level; it is therefore robust to small perturbations of
$\widehat{H}_{n}(1)$, which produce primarily a level shift in
$\widehat{H}_{n}(q)$.

\subsubsection{Stage 2: source of multiscaling}
\label{sssec:stage2}

If $\mathcal{H}^{(1)}_{0}$ is rejected, we test whether the observed
multiscaling can be reproduced by the marginal distribution of returns
alone. We formulate the null as
\begin{equation}
\mathcal{H}^{(2)}_{0}:\quad
\text{$\widehat{B}_{\text{rBergomi}}$ is consistent with what shuffled
surrogates produce.}
\end{equation}
A non-rejection of $\mathcal{H}^{(2)}_{0}$ indicates that no temporal
contribution to multiscaling can be detected at level $\alpha$, given
the marginal distribution of returns; it does not by itself prove that
multiscaling is purely distributional. A rejection indicates that
temporal dependence contributes detectably to $\widehat{B}$ beyond the
marginal distribution. The interpretation as ``distributionally
attributable'' versus ``temporally contributing'' is operational and
should be read accordingly.

We generate $J=1000$ shuffled surrogates and compute
$\widehat{B}_{\text{shuf},j}$ for each. Because the shuffled
distribution of $\widehat{B}$ can be skewed in heavy-tailed regimes
\citep{Barunik2010}, we use a robust distance-based two-sided statistic
\citep{Good2005,Phipson2010}. Let
\begin{equation}
\widetilde{B}\;=\;\mathrm{median}(\widehat{B}_{\text{shuf},1},\ldots,
\widehat{B}_{\text{shuf},J})
\end{equation}
denote the median of the surrogate distribution, more robust to outliers
than the mean. The signed distance of the original statistic from the
centre is
\begin{equation}
d_{\text{orig}}\;=\;|\widehat{B}_{\text{rBergomi}}-\widetilde{B}|,
\qquad
d_{\text{shuf},j}\;=\;|\widehat{B}_{\text{shuf},j}-\widetilde{B}|,
\end{equation}
and the two-sided $p$-value is
\begin{equation}
p_{\text{source}}\;=\;\frac{1}{J}\sum_{j=1}^{J}
\mathbb{I}\bigl(d_{\text{shuf},j}\geq d_{\text{orig}}\bigr).
\end{equation}
We reject $\mathcal{H}^{(2)}_{0}$ at level $\alpha=0.05$ if
$p_{\text{source}}<\alpha$. The two-sided form is appropriate because
temporal structure can in principle either enhance multiscaling (making
$\widehat{B}$ more negative than the shuffled distribution) or attenuate
it (less negative). If $\mathcal{H}^{(2)}_{0}$ is rejected with
$\widehat{B}_{\text{rBergomi}}<\widetilde{B}$, temporal dependence
\emph{enhances} multiscaling; if rejected with
$\widehat{B}_{\text{rBergomi}}>\widetilde{B}$, it \emph{attenuates} it.

\subsection{Implementation details}
\label{ssec:impl}

Throughout the paper we use $M=J=1000$ surrogates for both stages,
$R=1000$ Monte Carlo replicates of length $N=10{,}000$ per parameter
value, the adaptive choice of $q_{\max}$ described in
Section~\ref{ssec:hyper} with safety factor $s=0.8$, and the
fit-quality-driven choice of $\tau_{\max}$ with threshold
$R^{2}_{\min}\geq 0.95$. The multiscaling proxy $\widehat{B}$ is
estimated by the WLS regression~\eqref{eq:WLS}. We report in
Section~\ref{ssec:rb_results} the median and inter-quartile range of the
adaptively selected $q_{\max}$ and $\tau_{\max}$ across the 1000
simulations at each $H$ value.

\section{Results}
\label{sec:results}

We present our results in two parts. We first apply the methodology to
two synthetic processes with known properties, the Multifractal Random
Walk (MRW) and Fractional L\'evy Stable Motion (FLSM), to validate the
testing framework on cases where the source of any detected multiscaling
is known by construction (Section~\ref{ssec:synth}). We then apply the
validated methodology to the rough Bergomi model across a range of
roughness parameters (Section~\ref{ssec:rb_results}). For each parameter
value, we generate $R=1000$ independent Monte Carlo replicates of length
$N=10{,}000$ to ensure robust inference.

\subsection{Validation on synthetic models}
\label{ssec:synth}

To validate the methodology we apply it to two synthetic processes whose
multiscaling properties are known by construction. The first, MRW, is
\emph{designed} to be multifractal: any detection in the multifractal
regime should be attributed primarily to temporal dependence.\footnote{The
process also produces fat tails, so both effects coexist; the temporal
component dominates at larger values of the multifractality parameter
$\lambda$.} The second, FLSM, is \emph{designed} to combine heavy-tailed
innovations with fractional integration: when $H=0.5$ the
temporal-dependence component vanishes and FLSM reduces to an
$\alpha$-stable L\'evy process, i.e.~a purely heavy-tailed process with
no temporal dependence, so any detected multiscaling can come only from
the marginal distribution.

\subsubsection{Multifractal Random Walk}
\label{sssec:mrw}

The Multifractal Random Walk
\citep{bacry2001multifractal,BacryDelourMuzy2001PhysicaA,
MuzyBacry2002,Buonocore2017,brandi2022statistics} is a Gaussian process
with multiplicative log-volatility cascade: increments are of the form
$r_{t}=\varepsilon_{t}\exp(\omega_{t})$, where $\varepsilon_{t}$ is
i.i.d.~Gaussian and $\omega_{t}$ is a long-range correlated Gaussian
field whose covariance decays as
$\mathrm{Cov}(\omega_{t},\omega_{s})\sim\lambda^{2}\log(L/|t-s|)$ up to a
correlation length $L$. The parameter $\lambda^{2}$ controls the
strength of the multifractal cascade: $\lambda=0$ gives uncorrelated
Gaussian noise (uniscaling), while $\lambda>0$ produces genuine
multifractal scaling, with intermittency increasing in $\lambda$. The
source of multifractality in MRW is the long-range correlation of
$\omega$, i.e.~it is fundamentally \emph{temporal} in origin.

Table~\ref{tab:mrw_results} reports the application of our methodology
to MRW simulations with different $\lambda$ values. For small $\lambda$
the multifractal cascade is weak, the marginal distribution is
approximately Gaussian, and the Stage 1 detection rate is correspondingly
low (only $18.8\%$ of simulations at $\lambda=0.05$). The handful of
detected cases at low $\lambda$ should therefore be interpreted with
caution: they represent the right tail of the noise distribution under
near-Gaussianity, and the high distributional attribution
($\sim 95\%$) in this regime reflects the absence of strong cascade
structure rather than a genuine distributional source. As $\lambda$
grows, both the detection rate and the temporal attribution increase
sharply: at $\lambda=0.25$, the detection rate is $97.7\%$ and $78.5\%$
of the detected cases are attributed to temporal dependence, in
agreement with the known cascade-driven origin of multifractality in
MRW.

\begin{table}[H]
    \centering
    \caption{MRW Model: Multiscaling Significance and Source Attribution}
    \label{tab:mrw_results}
    \small
    \begin{tabular}{lccccc}
        \toprule
        \multirow{2}{*}{} & Multiscaling & \multicolumn{2}{c}{Source of Multiscaling} & \multicolumn{2}{c}{Multiscaling Statistics} \\
        \cmidrule(lr){2-2} \cmidrule(lr){3-4} \cmidrule(lr){5-6}
        $\lambda$ & Sig (\%) & Distributional(\%) & Temporal(\%) & Mean $\widehat{B}$ & SD($\widehat{B}$) \\
        \midrule
        0.050 & 18.8 & 95.2 & 4.8 & $-0.0018$ & 0.0039 \\
        0.150 & 78.5 & 75.3 & 24.7 & $-0.0094$ & 0.0053 \\
        0.250 & 97.7 & 21.5 & 78.5 & $-0.0248$ & 0.0081 \\
        \bottomrule
    \end{tabular}
\end{table}

The MRW validation establishes that, in regimes where genuine temporal
multifractal structure is present (large $\lambda$), our framework
correctly attributes the multiscaling to temporal dependence.

\subsubsection{Fractional L\'evy Stable Motion}
\label{sssec:flsm}

Fractional L\'evy Stable Motion (FLSM) is the fractional integration of
$\alpha$-stable noise with stability parameter $\alpha\in(0,2]$ and
Hurst parameter $H\in(0,1)$
\citep{huillet1999fractional,MazurOtryakhinPodolskij2020,
StoevTaqqu2004}. It combines two ingredients in a controlled way: heavy
tails (through the stable innovations) and long-range dependence
(through the fractional integration). FLSM is \emph{not} inherently
multiscaling; any apparent multiscaling detected on FLSM data must
therefore come either from finite-sample effects of the heavy tails or
from a non-trivial interaction with the fractional integration.

Two limit cases are particularly informative. At $\alpha=2$, FLSM
reduces to fractional Brownian motion (Gaussian, fractional, no heavy
tails); detected multiscaling should be negligible. At $H=0.5$, FLSM
reduces to pure $\alpha$-stable L\'evy motion (heavy tails, no
fractional dependence); this is a non-fractional process in which any
detected multiscaling can only have a distributional origin. The
$H=0.5$ row of Table~\ref{tab:flsm_results} therefore isolates the
non-fractional case: any multiscaling detected there, being an artefact
of the heavy tails rather than a genuine scaling property, should be
attributed to the marginal distribution.

Table~\ref{tab:flsm_results} reports the application of our methodology
to FLSM simulations with $\alpha=1.90$ and a range of Hurst values.

\begin{table}[H]
    \centering
    \caption{FLSM Model ($\alpha=1.90$): Multiscaling Significance and
    Source Attribution}
    \label{tab:flsm_results}
    \small
    \begin{tabular}{lccccc}
        \toprule
        \multirow{2}{*}{} & Multiscaling & \multicolumn{2}{c}{Source of Multiscaling} & \multicolumn{2}{c}{Multiscaling Statistics} \\
        \cmidrule(lr){2-2} \cmidrule(lr){3-4} \cmidrule(lr){5-6}
        H & Sig (\%) & Distributional(\%) & Temporal(\%) & Mean $\widehat{B}$ & SD($\widehat{B}$) \\
        \midrule
        0.10 & 23.0 & 93.9 & 6.1 & $-0.0003$ & 0.0058 \\
        0.50 & 23.3 & 94.0 & 6.0 & $-0.0020$ & 0.0078 \\
        0.90 & 27.3 & 65.2 & 34.8 & $-0.0107$ & 0.0103 \\
        \bottomrule
    \end{tabular}
\end{table}

Across the FLSM grid the detection rates are modest ($23$--$27\%$), as
expected given that the process is not inherently multiscaling: the
detections reflect finite-sample artefacts driven by heavy-tailed
innovations. For low and intermediate Hurst values ($H\in\{0.1,0.5\}$),
attribution is overwhelmingly distributional ($\sim 94\%$), confirming
that the framework correctly identifies the marginal distribution as
the source of apparent multiscaling. At $H=0.9$ the temporal
attribution rises to $34.8\%$, reflecting the contribution of
fractional integration in this regime; this is the most ambiguous case
and we therefore describe the FLSM family as primarily distributional
in the empirically relevant range of tail indices, rather than as
purely distributional. The two benchmarks play different and complementary roles. MRW at
$\lambda=0.25$ is a positive control: the multiscaling is genuine and
temporal by construction, and the procedure attributes it to temporal
dependence in $78.5\%$ of detected cases. FLSM at $H=0.5$ is a
specificity check rather than a positive control. The process has no
genuine multiscaling, the detection rate of $23.3\%$ is consistent with
the nominal size of the Stage~1 test inflated by heavy-tailed
finite-sample effects, and the informative content of the row is that
these rare detections are attributed to the marginal distribution in
$94\%$ of cases rather than being misread as temporal structure. Taken
together, the two benchmarks bracket the two ways the attribution could
fail: MRW shows that the procedure detects and correctly attributes
multiscaling that is genuinely temporal, and FLSM shows that it does not
manufacture spurious temporal attributions out of heavy tails alone. A
process with genuine multiscaling of distributional origin would be a
natural third case, and we return to this in
Section~\ref{sec:conclusion}.

\subsection{Main results: Rough Bergomi model}
\label{ssec:rb_results}

\subsubsection{Presence of multiscaling and selected hyperparameters}
\label{sssec:rb_presence}

\begin{figure}[H]
    \centering
\includegraphics[width=0.95\textwidth]{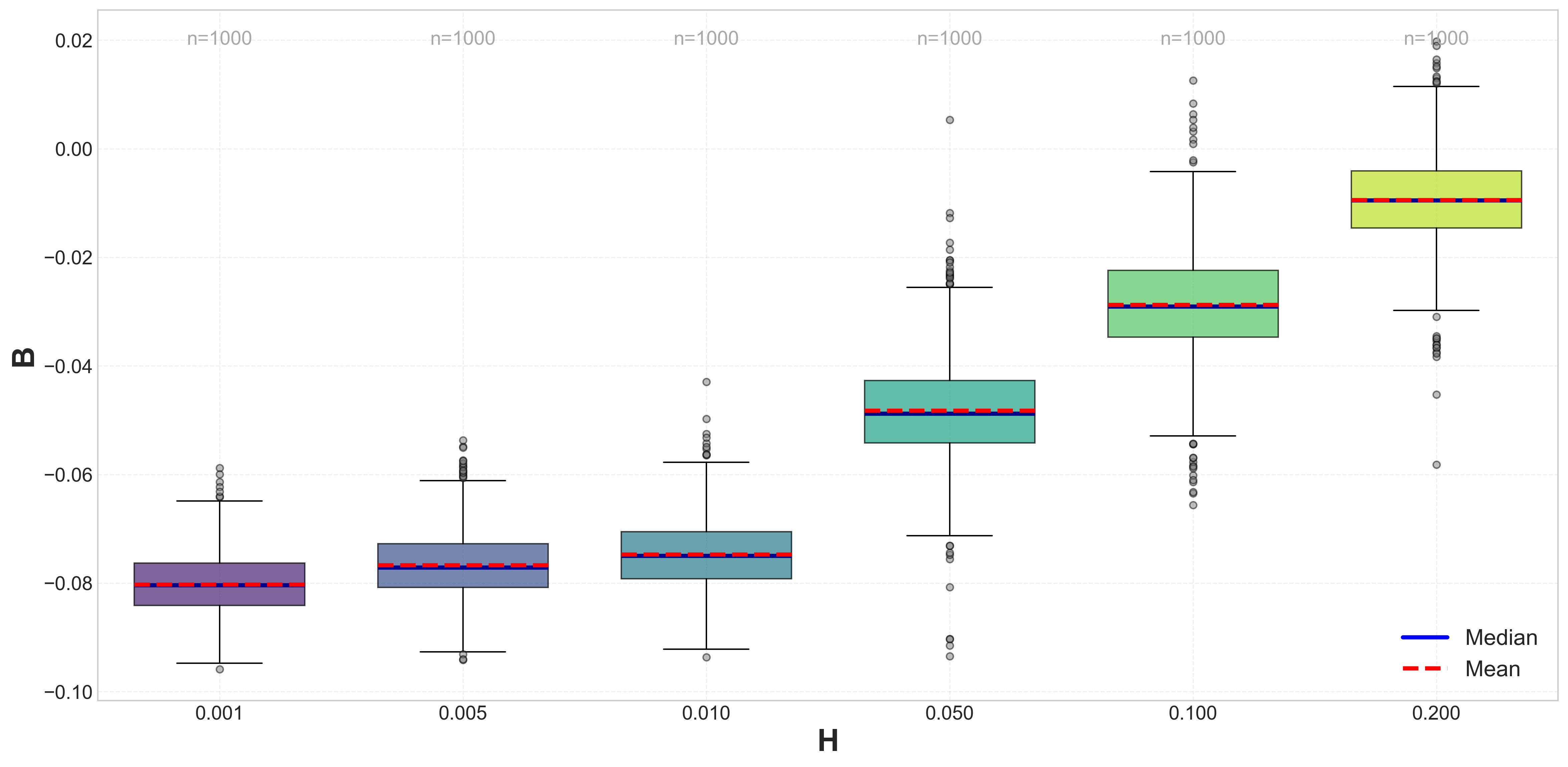}
    \caption{Multiscaling proxy $\widehat{B}$ as a function of $H$ in the
    rough Bergomi model. Boxplots over 1000 Monte Carlo replicates per
    $H$ value. As $H$ decreases, $\widehat{B}$ moves significantly away
    from zero, indicating progressively stronger multiscaling.}
    \label{fig:multiscaling_presence}
\end{figure}

We test the rBergomi model for the presence of multiscaling, against the
matched-fBm null, across the parameter grid $H\in[0.001,0.30]$.
Figure~\ref{fig:multiscaling_presence} reports the distribution of the
multiscaling proxy $\widehat{B}$ across $1000$ Monte Carlo replicates at
each $H$.

Before presenting the source attribution we comment briefly on the
operating range of the adaptive estimation procedure. For each Monte
Carlo replicate, $q_{\max}$ is set to $s\cdot\widehat{\alpha}_{\text{stable}}$
with $s=0.8$ and $\widehat{\alpha}_{\text{stable}}$ the maximum-likelihood
$\alpha$-stable index fitted to the simulated returns, and $\tau_{\max}$
is chosen by the fit-quality criterion~\eqref{eq:taustar} with threshold
$R^{2}_{\min}\geq 0.95$. By construction, $q_{\max}$ moves with the tail
heaviness of the simulated returns: it is smaller in the very-rough
regime (where the tails are heavier and fewer moments are reliably
finite) and larger in the moderate-roughness regime (where the tails
are lighter); in the rBergomi grid considered here, $q_{\max}$ remains
within the range $[1,2)$ throughout. The procedure is fully automated
and reproducible from the code accompanying this paper; the specific
selected values for any given simulation can be recovered by re-running
the publicly available code.

We emphasise that the adaptive $q_{\max}$ remains in a range where the
corresponding moments of the return distribution are theoretically
finite, given the fitted tail index. As noted in Section~\ref{ssec:hyper},
this is a deliberate methodological choice: strong curvature of
$\widehat{H}_{n}(q)$ at high $q$ in heavy-tailed regimes is a known
finite-sample artefact and would not reflect genuine multiscaling.
Restricting the analysis to the finite-moment range is therefore a
safeguard against spurious multiscaling detection, not a limitation of
statistical power. A useful consequence of this conservatism is that
our central finding is robust to the precise value of the safety
factor $s$ in~\eqref{eq:alpha_safe}: reducing $s$ further would shrink
the regression range in $q$ and only weaken the multiscaling signal
that the WLS regression can pick up; yet, as
Table~\ref{tab:rbergomi_very_rough} shows, multiscaling is detected in
$100\%$ of simulations at $H\leq 0.01$. \footnote{The strong-multiscaling
conclusion in the very-rough regime is therefore not an artefact of the
particular value $s=0.8$ but is robust to plausible variations of the
safety factor.}

Tables~\ref{tab:rbergomi_very_rough} and \ref{tab:rbergomi_moderate_rough}
report the source attribution. Detection rates are high or very high
across the rough regime ($H\leq 0.10$), and decrease to about $70\%$ at
$H=0.20$ where the multiscaling signal is genuinely weak.

\begin{table}[H]
    \centering
    \caption{rBergomi model: multiscaling significance and source
    attribution in the very-rough regime.}
    \label{tab:rbergomi_very_rough}
    \small
    \begin{tabular}{lccccc}
        \toprule
        \multirow{2}{*}{} & Multiscaling & \multicolumn{2}{c}{Source of Multiscaling} & \multicolumn{2}{c}{Multiscaling Statistics} \\
        \cmidrule(lr){2-2} \cmidrule(lr){3-4} \cmidrule(lr){5-6}
        H & Sig (\%) & Distributional(\%) & Temporal(\%) & Mean $\widehat{B}$ & SD($\widehat{B}$) \\
        \midrule
        0.001 & 100.0 & 95.3 & 4.7 & $-0.0802$ & 0.0058 \\
        0.005 & 100.0 & 90.3 & 9.7 & $-0.0767$ & 0.0064 \\
        0.010 & 100.0 & 78.4 & 21.6 & $-0.0747$ & 0.0068 \\
        \bottomrule
    \end{tabular}
\end{table}

\begin{table}[H]
    \centering
    \caption{rBergomi model: multiscaling significance and source
    attribution in the moderate-roughness regime.}
    \label{tab:rbergomi_moderate_rough}
    \small
    \begin{tabular}{lccccc}
        \toprule
        \multirow{2}{*}{} & Multiscaling & \multicolumn{2}{c}{Source of Multiscaling} & \multicolumn{2}{c}{Multiscaling Statistics} \\
        \cmidrule(lr){2-2} \cmidrule(lr){3-4} \cmidrule(lr){5-6}
        H & Sig (\%) & Distributional(\%) & Temporal(\%) & Mean $\widehat{B}$ & SD($\widehat{B}$) \\
        \midrule
        0.050 & 99.9 & 29.1 & 70.9 & $-0.0482$ & 0.0102 \\
        0.100 & 98.6 & 10.0 & 90.0 & $-0.0287$ & 0.0101 \\
        0.200 & 71.5 & 3.5 & 96.5 & $-0.0095$ & 0.0087 \\
        \bottomrule
    \end{tabular}
\end{table}

The pattern revealed by Tables~\ref{tab:rbergomi_very_rough} and
\ref{tab:rbergomi_moderate_rough} is striking and warrants careful
discussion. We observe two qualitatively different multiscaling regimes:

\begin{itemize}[leftmargin=*,itemsep=2pt]
  \item \textbf{Strong, tail-driven multiscaling at $H\to 0$.} The very
        rough regime exhibits very strong multiscaling
        ($|\widehat{B}|\approx 0.08$), detected in essentially every
        Monte Carlo replicate, and predominantly attributable to the
        marginal distribution of returns (78--95\% distributional).
        This is the regime in which the theoretical small-$H$
        asymptotics of \citet{forde2022riemann} apply: the integrated
        variance becomes extremely heavy-tailed in the rough limit,
        and this heavy tail propagates to the return distribution,
        driving the multiscaling proxy.
  \item \textbf{Mild, clustering-driven multiscaling at moderate $H$.}
        At $H=0.20$, the multiscaling is much weaker
        ($|\widehat{B}|\approx 0.01$), detected in only $\sim 70\%$ of
        replicates, and now predominantly attributable to temporal
        dependence ($96.5\%$ temporal). This is a qualitatively
        different phenomenon from the very-rough regime: the tails of
        the return distribution are much lighter at $H=0.2$ (see
        Section~\ref{sssec:rb_mech} below), and what remains of the
        multiscaling signal is genuinely associated with the volatility
        clustering that the model exhibits in this regime.
\end{itemize}

The detection of mild multiscaling at $H=0.20$ is not in conflict with
the theoretical result that strong tail-driven multiscaling in rBergomi
emerges only in the limit $H\to 0$. We are observing two distinct
mechanisms that produce multiscaling-like behaviour through the
$\widehat{B}$ statistic, the tail-driven mechanism that dominates as
$H\to 0$, and a much weaker, clustering-driven mechanism that survives
into the moderate-roughness regime. The transition between the two
occurs around $H\approx 0.05$--$0.10$.

\subsubsection{Distributional and temporal characteristics}
\label{sssec:rb_mech}

\begin{figure}[H]
    \centering
    \includegraphics[width=0.9\textwidth]{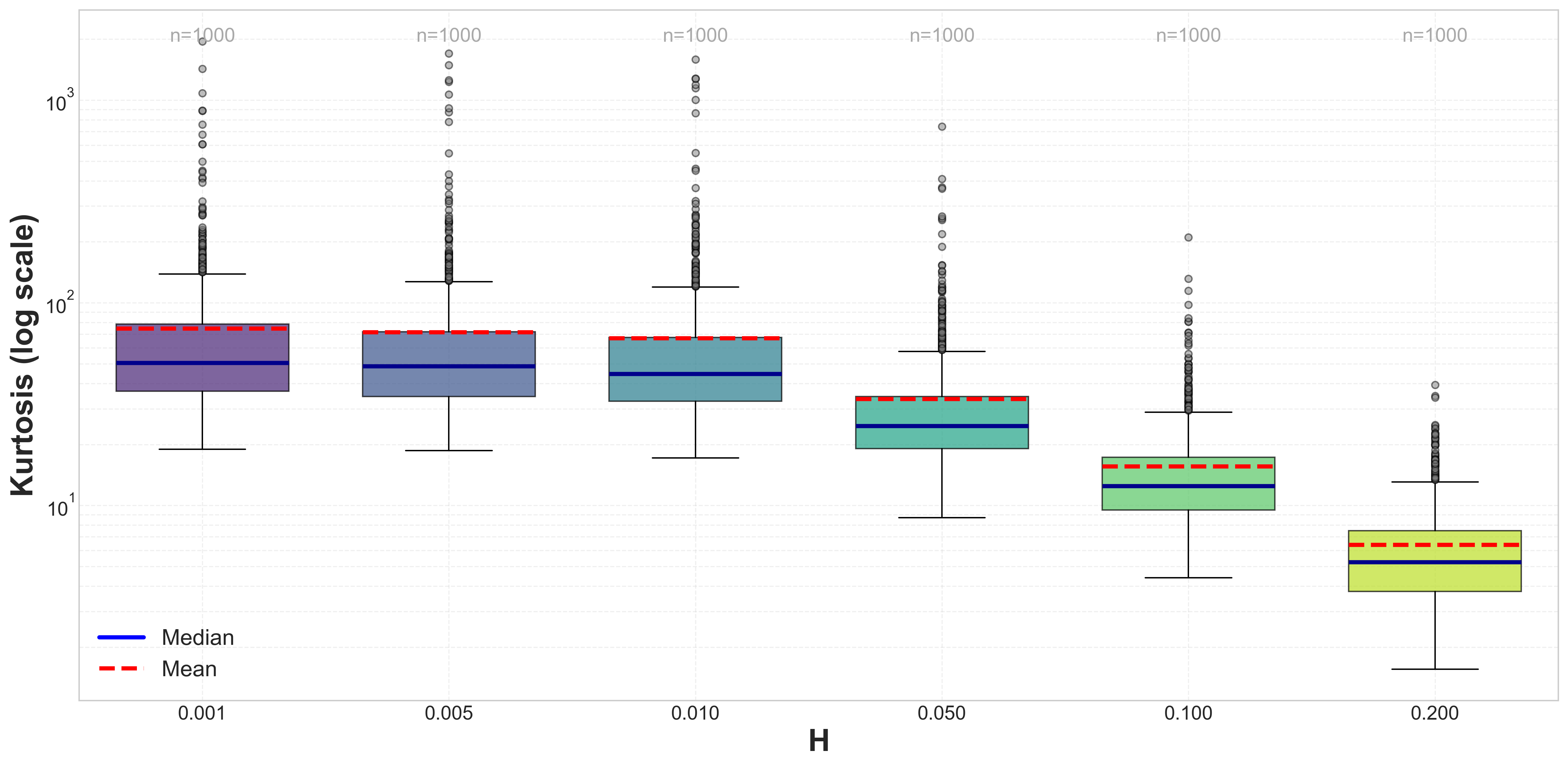}
    \caption{Kurtosis of returns (log scale) in the rough Bergomi model
    as a function of $H$. Tail heaviness decreases substantially as $H$
    increases, with median values dropping from $\sim 50$ at $H=0.001$
    to $\sim 5$ at $H=0.2$. Boxplots over 1000 Monte Carlo replicates.}
    \label{fig:kurtosis}
\end{figure}

\begin{figure}[H]
    \centering
    \includegraphics[width=0.9\textwidth]{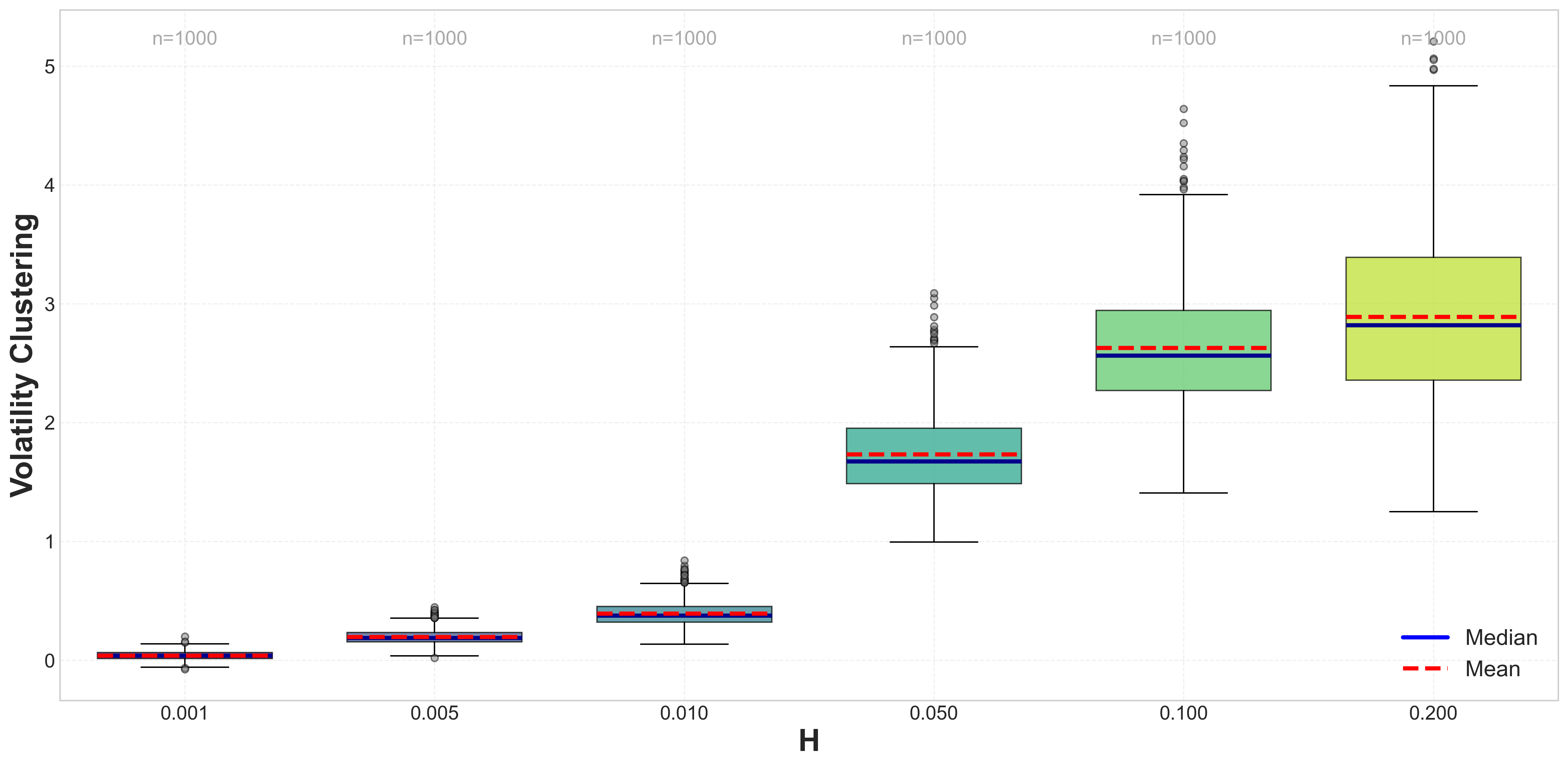}
    \caption{Volatility clustering in the rough Bergomi model as a
    function of $H$, measured as the sum of the first ten
    autocorrelations of absolute returns. Temporal persistence at lags
    1--10 increases substantially with $H$, rising from values close to
    zero at $H\leq 0.01$ to $\sim 3$ at $H=0.2$. Boxplots over 1000
    Monte Carlo replicates.}
    \label{fig:volclustering}
\end{figure}

The mechanism behind the regime transition can be visualised directly
by examining two summary statistics of the simulated paths across the
$H$ grid: the kurtosis of returns (capturing the tail heaviness of the
marginal distribution) and the sum of the first ten autocorrelations of
absolute returns (capturing volatility clustering). The two statistics
are reported in Figures~\ref{fig:kurtosis} and~\ref{fig:volclustering}.

The two figures provide a clean mechanistic explanation for the regime
transition. Figure~\ref{fig:kurtosis} shows that kurtosis decreases by
roughly an order of magnitude as $H$ increases from $0.001$ to $0.200$:
the marginal tails become substantially lighter as the volatility
becomes less rough, removing the main driver of the tail-attributed
multiscaling. Conversely, Figure~\ref{fig:volclustering} shows that
volatility clustering at short lags, nearly absent in the very-rough
regime, increases substantially as $H$ rises towards $0.20$,
providing a new (and quantitatively weaker) source of multiscaling.

A clarification regarding the very-rough regime is in order. fBm with
$H\to 0$ is strongly anti-persistent in its increments; what
Figure~\ref{fig:volclustering} reports, however, is the autocorrelation
of absolute returns, which measures volatility clustering rather
than directional persistence. The statement that ``temporal dependence
contributes minimally to multiscaling as $H\to 0$'' refers specifically
to this volatility-clustering component, which is the relevant temporal
structure for the source attribution: the anti-persistence of the
underlying fractional driver does not translate into clustering of
squared/absolute returns in this regime.

Taken together, our results suggest a re-reading of the rBergomi model's
multiscaling. The dramatic multiscaling visible at small $H$ is
predominantly inherited from the heavy-tailed marginal distribution that
the rough variance process generates; at moderate $H$, the marginal
tails are not heavy enough to dominate and what remains of the
multiscaling signal is genuinely associated with the volatility
clustering that emerges in this regime. Both mechanisms can produce
multiscaling through the $\widehat{B}$ statistic, but they are
operationally and quantitatively distinct.

This combination is empirically interesting: at intermediate $H$, the
model can exhibit volatility clustering together with mild multiscaling,
a combination consistent with stylised facts of real financial assets
in which both persistent volatility dynamics and moderate departures
from simple scaling are documented \citep{cont2001empirical,di2007multi}.

\section{Discussion and Conclusions}
\label{sec:conclusion}

This paper has investigated the source of multiscaling behaviour in the
rough Bergomi model, using a two-stage statistical testing procedure
based on matched-fBm and shuffled surrogates. Our principal finding is
that the multiscaling exhibited by the rBergomi model has two distinct
sources, operating in two different regions of the roughness parameter
space.

In the very-rough regime ($H\leq 0.01$), the multiscaling is strong in
magnitude and predominantly attributable to the marginal distribution
of returns: the integrated variance becomes extremely heavy-tailed as
$H\to 0$, and this heavy tail propagates to the returns and drives the
scaling of high-order moments. Between $78\%$ and $95\%$ of the detected
multiscaling in this regime is attributable to the marginal
distribution. This is consistent with, and gives an empirical
interpretation of, the small-$H$ asymptotics of
\citet{forde2022riemann}, and it confirms the multiscaling behaviour
documented in our earlier work \citep{brandi2022multiscaling} while
reinterpreting it as a distributional rather than a temporal
phenomenon.

In the moderate-roughness regime ($H\geq 0.10$), the picture changes
qualitatively. The multiscaling is weaker in magnitude, detected in a
smaller fraction of simulations (about $70\%$ at $H=0.20$), and is
predominantly attributable to temporal dependence rather than to the
marginal distribution. This is not a contradiction with the
distributional finding in the very-rough regime: the two regimes
correspond to two different mechanisms by which the rBergomi model can
generate multiscaling. We document this transition mechanistically by
showing that the kurtosis of returns decreases by an order of magnitude
between $H=0.001$ and $H=0.20$ while the autocorrelation of absolute
returns at short lags grows correspondingly. The crossover occurs
around $H\approx 0.05$--$0.10$.

The methodological contribution of the paper is the two-stage
surrogate-based framework itself. The combination of matched-fBm
surrogates for testing the presence of multiscaling and shuffled
surrogates for assessing whether temporal dependence contributes beyond
the marginal distribution provides a more rigorous alternative to the
classical $t$-test against $B=0$, which ignores the finite-sample
distribution of the multiscaling proxy under uniscaling. The
distance-based two-sided statistic in the second stage handles the
asymmetry of the null distribution that arises in heavy-tailed regimes.
The framework was validated on two synthetic processes for which the
source of any detected multiscaling is known by construction: the
Multifractal Random Walk, where the multifractality is built from a
log-volatility cascade and our framework correctly attributes the
detected multiscaling to temporal dependence, and Fractional L\'evy
Stable Motion, including the limit case $H=0.5$ in which FLSM reduces
to a non-fractional $\alpha$-stable L\'evy process whose apparent
multiscaling can only have a distributional origin.

The results have implications for the way the rBergomi model is
interpreted in financial applications. The model's option-pricing value
is unaffected by our findings, since the relevant calibrated
$H\in[0.05,0.15]$ is precisely the range in which the model generates
volatility clustering and mild multiscaling consistent with empirical
stylised facts \citep{cont2001empirical}. From the point of view of
risk management and scaling-based applications, however, our results
suggest that the rBergomi model captures the empirical multiscaling of
financial returns through two different channels, and that the dominant
channel depends on the calibrated value of $H$. At calibrations close
to the empirically reported range, the temporal channel dominates,
which is consistent with the conventional reading of financial
multiscaling as reflecting volatility persistence; in extrapolation
towards $H\to 0$, the distributional channel becomes dominant. Risk
applications that rely on extrapolating scaling laws to long horizons
should be aware of this distinction.

Several directions for future work are suggested by our analysis.
First, the application of the same methodology to other rough
volatility models, rough Heston, the quadratic rough Heston model
\citep{gatheral2020quadratic}, and multivariate rough volatility models
with asset-specific Hurst exponents \citep{dugogiorgio2026}, would
clarify whether the regime structure documented here is specific to
rBergomi or shared across the rough volatility family. Our expectation
is that the qualitative regime transition (heavy tails dominating as
$H\to 0$, clustering dominating at intermediate $H$) is generic, but
quantitative attribution will depend on the specifics of each variance
process. The methodology could similarly be applied to non-fractional
multiscaling models. The class introduced by
\citet{daipra2015multiscaling} is a particularly interesting case: there
the volatility follows a mean-reverting equation driven by a L\'evy
subordinator whose characteristic measure has power-law tails, and the
absolute moments of the increments scale as $h^{q/2}$ below a threshold
moment order $q^{*}$ and as $h^{A(q)}$ with $A(q)<q/2$ above it, with
multiscaling occurring precisely when the mean reversion is superlinear.
This is genuine multiscaling, analytically characterised, generated by
the tails of the volatility process rather than by fractional memory.
Applying our two-stage procedure there would test the attribution
machinery against a distributional mechanism of a quite different kind
from the one operating in rBergomi, complementing the cascade-driven
case supplied by MRW, and we regard it as the most informative single
extension of the validation programme. Second, empirical calibrations of the rBergomi model to option
data typically identify a one-dimensional ridge of nearly equivalent
$(H,\eta)$ combinations in parameter space rather than a single
well-identified point \citep{bayer2016pricing,romer2022empirical}: lower
values of $H$ tend to be compensated by different values of $\eta$ along
an iso-fit curve. Varying $H$ and $\eta$ jointly along this ridge,
rather than holding $\eta$ fixed as in the present analysis, would
provide a more representative picture of the model's scaling behaviour
over the empirically relevant region of parameter space. Third, an empirical application of the methodology to real
financial data would close the loop with the original motivation for
rough volatility models, by quantifying the relative contributions of
distributional and temporal sources to the multiscaling observed in
financial returns themselves. We leave these directions to future work.

In conclusion, the apparent complexity of the rBergomi model, as
manifested through multiscaling, has two distinct sources rather than
one: a tail-driven mechanism that dominates in the very-rough limit and
a clustering-driven mechanism that survives into the moderate-roughness
regime. Distinguishing between them is methodologically important and,
we believe, also conceptually relevant for the interpretation of rough
volatility models in financial applications.


\section*{Acknowledgements}
We thank two anonymous referees for their careful reading and insightful comments, which have significantly improved the paper. We are also grateful to Pasquale Casaburi for valuable discussions in the early stages of this project.
\section*{Declaration of competing interests}
The authors declare that they have no known competing financial interests or personal relationships
that could have appeared to influence the work reported in this paper.
\section*{Declaration of funding}
No funding was received.

\bibliographystyle{rQUF}
\bibliography{biblio}

\end{document}